 \documentclass[aps,pra,twocolumn,superscriptaddress,amsmath,amssymb,showpacs]{revtex4}
\usepackage{graphicx}
\usepackage{amsmath}
\usepackage{amssymb}
\usepackage{mathrsfs}
\usepackage{amsfonts}
\usepackage{dcolumn}
\usepackage{bm}

\newcommand{\vvr}{\mathbf{r}}
\newcommand{\vx}{\mathbf{x}}

\newcommand{\vp}{\mathbf{p}}
\newcommand{\vq}{\mathbf{k}_\perp}
\newcommand{\vk}{\mathbf{k}}

\newcommand{\vE}{\mathbf{E}}

\newcommand{\hvE}{\hat{\mathbf{E}}}
\newcommand{\hvB}{\hat{\mathbf{B}}}

\newcommand{\hvP}{\hat{\mathbf{P}}}
\newcommand{\hvJ}{\hat{\mathbf{J}}}
\newcommand{\hvS}{\hat{\mathbf{S}}}

\newcommand{\hvx}{\hat{\mathbf{x}}}
\newcommand{\hvy}{\hat{\mathbf{y}}}
\newcommand{\hvz}{\hat{\mathbf{z}}}
\newcommand{\ha}{\hat{a}}

\newcommand{\he}{\hat{\bm{e}}}
\newcommand{\hx}{\hat{\bm{x}}}
\newcommand{\hy}{\hat{\bm{y}}}
\newcommand{\hz}{\hat{\bm{z}}}
\newcommand{\hn}{\hat{\bm{n}}}

\newcommand{\hbx}{\hat{\bm{x}}}
\newcommand{\hby}{\hat{\bm{y}}}
\newcommand{\hbz}{\hat{\bm{z}}}
\newcommand{\kp}{k_\perp}

\providecommand{\abs}[1]{\lvert#1\rvert}

\newcommand{\lbar}{{\lambda \negthickspace \! \text{\rule[5.5pt]{0.135cm}{.2pt}} \, }}
\newcommand{\di}{\mathrm{d}}

\begin{document}
\title{Transverse angular momentum of photons}
\author{Andrea Aiello}
\email{andrea.aiello@mpl.mpg.de}
\affiliation{Max Planck Institute for the Science of Light, G\"{u}nter-Scharowsky-Stra{\ss}e 1/Bau 24, 91058 Erlangen, Germany.}
\author{Christoph Marquardt}
\affiliation{Max Planck Institute for the Science of Light, G\"{u}nter-Scharowsky-Stra{\ss}e 1/Bau 24, 91058 Erlangen, Germany.}
\affiliation{Institute for Optics, Information and Photonics, University Erlangen-N\"{u}rnberg,\\ Staudtstra{\ss}e 7/B2, 91058 Erlangen, Germany.}
\author{Gerd Leuchs}
\affiliation{Max Planck Institute for the Science of Light, G\"{u}nter-Scharowsky-Stra{\ss}e 1/Bau 24, 91058 Erlangen, Germany.}
\affiliation{Institute for Optics, Information and Photonics, University Erlangen-N\"{u}rnberg,\\ Staudtstra{\ss}e 7/B2, 91058 Erlangen, Germany.}
\begin{abstract}
We develop the quantum theory of transverse angular momentum of light beams. The theory applies to paraxial and quasi-paraxial photon beams in vacuum, and reproduces the known results for classical beams when applied to coherent states of the field. Both the Poynting vector, alias the linear momentum, and the angular momentum quantum operators of a light beam are calculated including contributions from first-order transverse derivatives. This permits a correct description of the energy flow in the beam and the natural emergence of both the spin and the angular momentum of the photons. We show that for collimated beams of light, orbital angular momentum operators do not satisfy the standard commutation rules. Finally, we discuss the application of our theory to some concrete cases.
\end{abstract}
\pacs{03.70.+k, 42.50.-p, 42.50.Tx} \maketitle
%
%
%
%
%
%
%
%
%
\section{introduction}
The quantum theory of light assigns  a longitudinal component ${\mathbf{J} \cdot \mathbf{k}/\abs{\mathbf{k}} = \hbar \sigma}$ of spin angular momentum (SAM) to a photon of energy $\hbar \omega$ and momentum $\hbar \mathbf{k}$, where $\sigma = \pm 1$ for a circularly polarized photon, and $\sigma = 0$ for a linearly polarized one.
At optical frequencies, however, the representation of a photon by a single plane wave mode of sharp angular frequency $\omega$ and wave vector $\vk$ is quite unrealistic. Rather, a bona fide optical photon should be described as a wave packet formed by the superposition of many (possibly infinite) plane waves of different frequencies and wave vectors.
As a result of this superposition a complex  spatial structure of the photon field may be generated and tailored in order to  carry an orbital angular momentum (OAM).
Such possibility was envisaged in 1992 by Allen, Woerdman and coworkers \cite{Allen92}  who showed that a Laguerre-Gauss beam of light \cite{SiegmanBook} propagating in the $z$ direction with a wave front of the form $(x + i y \ell/\abs{\ell})^{\abs{\ell}}$, possesses a $z$ component of orbital angular momentum of $\hbar \ell$ per photon \cite{NotePhoton}, with $\ell \in \{0 \pm1,\pm2,\ldots \}$. This result boosted the interest of the physics community
for light beams with angular momentum which had found numerous applications
  ranging from quantum cryptography \cite{PhysRevA.68.012323} to the realization of EPR entangled systems \cite{hsu:043825} (see, e.g., \cite{BarnettBook} and \cite{FrankeA} for recent surveys).

Previous authors have presented classical \cite{vanEnk&Nienhuis,Barnett&Allen} and quantum \cite{vanEnk&Nienhuis94,calvo:013805} treatments of light beams with spin and orbital angular momentum. In these studies the attention was mainly devoted to the longitudinal (namely parallel to the beam propagation direction) component of the angular momentum. This was probably due to the fact that the spin angular momentum of  photons can only be defined along the direction of propagation. However, it was very recently noticed that transverse, as opposed to longitudinal, components of angular momentum may be responsible for interesting phenomena as, e.g., the so-called geometric spin Hall effect of light \cite{aiello:100401,Bekshaev:09}. A classical theory of transverse optical angular momentum was developed in \cite{aiello:100401}.

In this paper we present a quantum theory of transverse angular momentum of photons. We apply the exact quantization scheme for light beams described in \cite{AielloPRAParax,AielloOLParax}, to the development of a rigorous theory of quasi-paraxial photon fields, namely fields represented by the Lax \textit{et al.}  power series expansion \cite{PhysRevA.11.1365} truncated at first-order terms. The zero-order terms in the Lax expansion are exact solutions of the paraxial wave equation. However, it was shown in \cite{HausandPan,EandS} that the presence of these terms solely is not enough to guarantee a correct description of both the energy flow and the spin angular momentum in the beam. Thus, we have included first-order transverse derivatives in our description of the photon fields. In this manner we were able to build a self-consistent theory of transverse angular momentum which displays some nontrivial characteristics  as, e.g.,  anomalous commutation relations between angular momentum operators.

This paper is structured as follows: In Sec. II we shortly review the exact quantization scheme  \cite{AielloPRAParax,AielloOLParax} and apply it to the present scenario. Then, in Sec. III we derive closed expressions for both the linear and angular momentum operators of the photon fields. By using these results, we show in Sec. IV that such operators do not fulfill canonical commutation relations in the paraxial regime of propagation, and discuss these findings. Subsequently, in Sec. V, we study some specific states of the fields that illustrate the occurrence of transverse components of the angular momentum. Finally, in Sec. VI we summarize our results.
\section{Quantization of the fields}
In this section we illustrate the quantization procedure for quasi-paraxial beams of light.
\subsection{Exact field quantization}
In Ref.  \cite{AielloOLParax} it was demonstrated that for a light beam propagating in the forward $z$ direction,
the plane-wave expansion of the positive-frequency part of the electric field operator in the Coulomb gauge  can be written as:
\begin{widetext}
\begin{align}
\hvE^+(\vvr,t) =& \;  i \iint\limits_{-\infty}^{\quad  \infty} \di k_x \, \di k_y \,  \int_{c \kp}^\infty \di \omega  \left[\frac{\hbar \omega/ \kappa_z(\kp,\omega)}{16 \pi^3 \varepsilon_0 c}\right]^{1/2} \exp{\left[-i \omega \left( t- z /c\right) \right]} \nonumber \\
&  \times \sum_{\mu=1}^2 \he_\mu(\vq,\omega) \ha_\mu(\vq,\omega) \exp \left[i \vq \cdot \vx - i z \frac{\omega}{c} \bigl(1 - \kappa_z \left(\kp,\omega\right) \bigr) \right], \label{E}
\end{align}
\end{widetext}
where $\vvr = (x,y,z)$ is the position vector and  $\vx = (x,y)$ its transverse part in the plane $z=$ const. Moreover, $\vq = (k_x,k_y)$ is the transverse part of the wave vector  $\vk = \left(k_x,k_y, k_z\right)$, with $k \equiv \abs{\vk} = \omega/c$, and
\begin{align}
 \frac{k_z}{k} \equiv \kappa_z(\kp,\omega)= \left(1 - \frac{\kp^2 c^2}{\omega^2} \right)^{1/2}, \label{alpha}
\end{align}
where  $\kp =\abs{\vq} =(k_x^2 + k_y^2)^{1/2}$. Note that  the function $\kappa_z(\kp,\omega)\geq 0$ is strictly defined only for $\omega \geq \kp c$, and  that
only the parts of the ¯field propagating in the positive $z$
direction are included in Eq. (\ref{E}).

 The  operator $\ha_\mu(\vq,\omega)$ annihilates a photon with wave vector $\vk$ and polarization $\he_\mu(\vq,\omega)$, and  satisfies the canonical commutation rules
\begin{align}
\bigl[ \ha_\mu(\vq,\omega),\ha_{\mu'}^\dagger(\vq',\omega')\bigr] = \delta_{\mu \mu'} \delta(\vq - \vq') \delta(\omega -\omega'). \label{CCR}
\end{align}
The three  unit vectors $\bigl\{\he_1(\vq,\omega),\he_2(\vq,\omega),  \vk/k \bigr\}$ form a right-handed Cartesian frame that can be obtained from the reference basis $\{\hx, \hy,\hz \}$ via a rotation around the axis
\begin{align}
\hn= {\hz \times \vk}/{\abs{\hz \times \vk}}, \label{hn}
\end{align}
by the angle $\theta$  between  $\vk$ and  $\hz$:
\begin{align}
\theta = \arcsin(\kp/k) = \arcsin(\kp c/\omega). \label{rotAngle}
\end{align}
If we represent such a rotation  by means of the Rodrigues' formula \cite{Rodrigues} via  the matrix
\begin{align}
R \left( \theta, \hn  \right) = \exp \left( \theta E \right) = I + E \sin \theta + E^2(1-\cos \theta), \label{rotation1}
\end{align}
where $I$ is the $3 \times 3$ identity matrix and  $E$ denotes the antisymmetric $3 \times 3$ matrix of elements
\begin{align}
E_{ij} = -\sum_{k=1}^3\varepsilon_{ijk} (\hn)_k , \qquad i,j \in \{ 1,2,3\}, \label{MatE}
\end{align}
($\varepsilon_{ijk}$ is the completely antisymmetric Levi-Civita symbol), then we obtain
\begin{subequations}\label{basis}
\begin{align}
\he_1 (\vq,\omega) \;= & \; R \left( \theta, \hn  \right) \hx ,\\
\he_2 (\vq,\omega)  \;= & \;  R \left( \theta, \hn  \right) \hy,
\end{align}
\end{subequations}
and $\vk/k  =  R \left( \theta, \hn  \right) \hz$. It is worth noting that from Eq. (\ref{rotAngle}) and the definition (\ref{alpha}), it follows that
\begin{align}
\sin \theta = \kp c/\omega \quad \Rightarrow \quad \kappa_z(\kp,\omega) = \cos \theta. \label{alphaCos}
\end{align}
The geometry of the ``global'' $\{\hx,\hy,\hz \}$ and the ``local'' $\bigl\{\he_1(\vq,\omega),\he_2(\vq,\omega),  \vk/k \bigr\}$ Cartesian frames, is illustrated in Fig. 1.
\begin{figure}[!h]
\includegraphics[angle=0,width=7truecm]{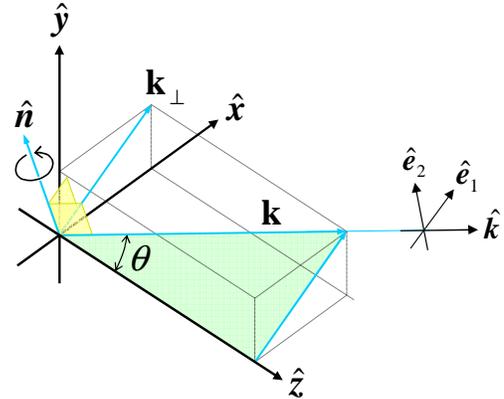}
\caption{\label{fig:1} Illustrating the geometry of the problem.}
\end{figure}
\subsection{Quasi-paraxial quantization}
Equation (\ref{E}) is exact, therefore it can be used to describe any field that propagates in the positive $z$ direction. However, most optical experiments use narrow-band and well-collimated beams which satisfy, respectively,  the conditions
\begin{align}
\Delta \omega \ll \omega_0, \qquad \text{and} \qquad  \theta_0 \ll 1, \label{Mono&Parax}
\end{align}
where $\omega_0$ is the central frequency of the bandwidth $\Delta \omega$, and $\theta_0$ is the angular spread of the beam around the central wave vector $\vk_0 = \hz \omega_0/c$.
For states of the radiation field
whose excitation bandwidths and angular apertures  satisfy Eq. (\ref{Mono&Parax}),
 the $\vk$-space  in Eq. (\ref{E}) can be restricted, without significant error, to the intersection $\mathcal{I}$ between the cone of axis $\vk_0 \propto \hz$  and aperture $2 \theta_0$, and the spherical shell of radius $\omega_0/c$ and thickness $\Delta \omega/c$, as shown in Fig. 2.
\begin{figure}[!h]
\includegraphics[angle=0,width=7truecm]{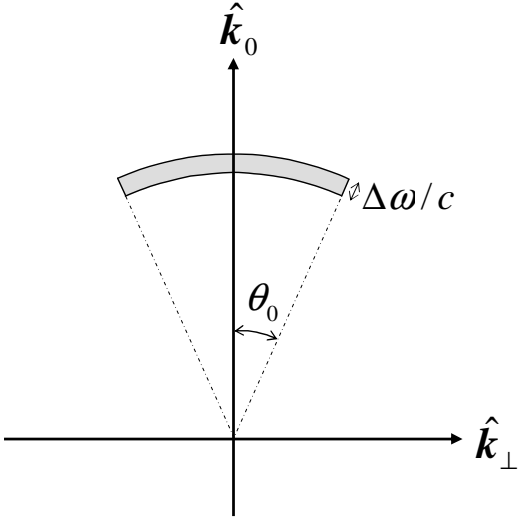}
\caption{\label{fig:2} The grey area represents the domain  $\mathcal{I}$ within the $\mathbf{k}$-space.}
\end{figure}
 For any $\vk \in \mathcal{I}$ we have $\theta \lesssim \theta_0 \ll 1$, and  from Eqs. (\ref{rotAngle},\ref{alphaCos}) it  follows that
\begin{align}
\kappa_z(\kp,\omega) \, \simeq  \, 1 - \frac{1}{2}\sin^2 \theta  \,  =  \, 1 - \frac{1}{2} \left({\kp c}/{ \omega}\right)^2, \label{kappa2}
\end{align}
where  $ \kp c/\omega  \ll 1$. Thus, the function $\kappa_z(\kp,\omega)$ inside the square root in Eq. (\ref{E}) can be approximated by $1$, while the second order term  must be retained in the $z$-dependent part of the last exponential in Eq. (\ref{E}) to obtain a nonzero result:
\begin{align}
- i z \frac{\omega}{c} \bigl(1 - \kappa_z\left(q,\omega\right) \bigr) \simeq  - i z \frac{\kp^2 c}{2 \, \omega} .\label{ExpApprox}
\end{align}
Moreover,  we can approximate Eq. (\ref{rotation1}) with
\begin{align}
R \left( \theta, \hn  \right) \simeq I + E \sin \theta = I + (\kp c/\omega) E, \label{rotation2}
\end{align}
and from Eq. (\ref{basis}) it readily follows that
\begin{subequations}\label{basisParax}
\begin{align}
\he_1 (\vq,\omega) \;\simeq  &  \; \hx - \hz k_x/k ,\\
\he_2 (\vq,\omega) \;\simeq  &  \;  \hy - \hz k_y /k .
\end{align}
\end{subequations}
Finally, for excitations of the field satisfying Eq. (\ref{Mono&Parax}), we can extend the  integration over $\omega$ in (\ref{E}) from $0$ to $\infty$ without relevant error, and the positive-frequency part of the electric field operator  can be written as
\begin{align}
\hvE^+(\vvr,t)= & \; \; i  \int_0^\infty \di \omega \left( {\frac{\hbar \omega}{16 \pi^3 \varepsilon_0 c}}\right)^{1/2} \exp \bigl[-i \omega(t-z/c) \bigr] \nonumber \\
& \, \times \int \di^2 k_\perp \sum_{\mu=1}^2 \left( \hat{\bm{x}}_\mu - \hz \frac{ c}{\omega}k_{x_\mu} \right) \ha_\mu(\vq,\omega) \nonumber \\
& \, \times\exp \left(i \vq \cdot \vx - i z \frac{\kp^2 c}{2 \, \omega} \right), \label{Eparax}
\end{align}
where $\di^2 k_\perp  = \di k_x \, \di k_y$, $x_1 = x,\; x_2 = y$, and we interchanged the order of integration. Equation (\ref{Eparax}) can be further simplified by noting that
\begin{align}
-k_{x_\mu}\exp \left(i \vq \cdot \vx \right) = i \frac{\partial}{\partial x_\mu}\exp \left(i \vq \cdot \vx \right), \label{Simplify}
\end{align}
which permits us to rewrite Eq. (\ref{Eparax}) as
\begin{align}
\hvE^+(\vvr,t)= & \, i  \int_0^\infty \di \omega \left( {\frac{\hbar \omega}{4 \pi \varepsilon_0 c}}\right)^{1/2} \exp \bigl[-i \omega(t-z/c) \bigr] \nonumber \\
& \, \times  \sum_{\mu=1}^2 \left( \hat{\bm{x}}_\mu +i \hz \frac{c}{\omega} \hx_\mu \cdot \bm{\nabla}_\perp \right)\ha_\mu(\vx,z,\omega), \label{Eparax2}
\end{align}
where we have defined $\bm{\nabla}_\perp = \hat{\bm{x}} \,\partial/ \partial x + \hat{\bm{y}} \, \partial/ \partial y$, and
\begin{align}
 \ha_\mu(\vx,z,\omega) = & \, (2 \pi)^{-1} \int {\di^2 k_\perp } \, \ha_\mu(\vq,\omega) \nonumber \\
& \, \times \exp \left( i \vq \cdot \vx - i z \frac{\kp^2 c}{2 \, \omega}  \right). \label{Aparax}
\end{align}
At any plane $z = \text{const.}$, this is still a bona fide quantum harmonic oscillator annihilation operator, as it satisfies the following commutation rules:
\begin{align}
\bigl[ \ha_\mu(\vx,z,\omega),\ha_{\mu'}^\dagger(\vx',z,\omega')\bigr]\! = \delta_{\mu \mu'} \delta(\vx - \vx') \delta(\omega -\omega'). \nonumber
\end{align}
The integral (\ref{Aparax}) can now be evaluated by using the following relation \cite{calvo:013805}
\begin{align}
 \frac{1}{2 \pi}\exp \biggl( i \vq \cdot \vx  \biggr. &  \biggl. - \, i z \frac{\kp^2 c}{2 \, \omega}  \biggr)  \nonumber \\
& \, =  \sum_{n,m} \widetilde{\psi}_{nm}^*(\vq, \omega) \, \psi_{nm} (\vx,z,\omega), \label{Calvo1}
\end{align}
where $\psi_{nm} (\vx,z,\omega)$ is a complete orthonormal set of functions  on $\vx$,
\begin{align}
 \int\psi_{nm}^* (\vx,z,\omega) \, \psi_{n' m'} (\vx,z,\omega) \, \di^2 x  \, = & \,\delta_{n n'}\delta_{mm'}, \nonumber \\
 \sum_{n, \, m} \psi_{nm}^* (\vx,z,\omega)\, \psi_{nm} (\vx ',z,\omega) \, = & \, \delta(\vx - \vx'), \label{Orthonormal}
\end{align}
and $n, m \in \mathbb{Z}$ are the appropriate integer labels for the set.
In order to fulfill Eq. (\ref{Calvo1}), the functions  $\psi_{nm} (\vx,z,\omega)$ must be chosen amongst either the Hermite-Gauss (HG) or the Laguerre-Gauss (LG) sets of solutions of the paraxial wave equation \cite{SiegmanBook}, and
\begin{align}
 \widetilde{\psi}_{nm}(\vq , \omega) = (2 \pi)^{-1} \int  \psi_{nm} (\vx,0,\omega) \exp(-i \vx \cdot \vq )\,\di^2 x, \label{FT}
\end{align}
is just the Fourier transform of $\left. \psi_{nm} (\vx,z,\omega) \right|_{z=0}$ evaluated at $z=0$.
Using (\ref{Calvo1}) inside (\ref{Aparax}) we obtain
\begin{align}
\ha_\mu(\vx,z,\omega) = \sum_{n,m}\ha_{\mu n m}(\omega)\psi_{nm}(\vx,z, \omega), \label{Aparax2}
\end{align}
where the operator
\begin{align}
\ha_{\mu nm}(\omega) =  \int \di^2 k_\perp  \,  \widetilde{\psi}_{nm}^*(\vq \,) \, \ha_\mu(\vq,\omega).
 \label{Aparax3}
\end{align}
annihilates a photon with polarization  $\hx_\mu$ in the spatial mode $\psi_{nm}$.
From Eq. (\ref{CCR}) and exploiting the orthogonality of the  modes $\psi_{nm}$, it is easy to verify that
\begin{align}
\bigl[\ha_{\mu nm}(\omega),\ha_{\mu' n' m'}^\dagger(\omega')\bigr] = \delta_{\mu \mu'} \delta_{nn'} \delta_{mm'} \delta(\omega -\omega'). \label{CCR2}
\end{align}
Finally, we can write the positive-frequency part of the electric field operator as
\begin{align}
\hvE^+ & \!(\vvr,t)=  i  \int_0^\infty \di \omega \left( {\frac{\hbar \omega}{4 \pi \varepsilon_0 c}}\right)^{1/2} \exp \bigl[-i \omega(t-z/c) \bigr] \nonumber \\
&  \times  \sum_{\mu,n,m} \ha_{\mu nm}(\omega)\left( \hat{\bm{x}}_\mu +i \hz \frac{c}{\omega} \hx_\mu \cdot \bm{\nabla}_\perp \right)\psi_{nm}(\vx,z, \omega). \label{Eparax3}
\end{align}
It should be noticed in the expression above the presence of the transverse gradient $\bm{\nabla}_\perp $ which is equivalent to the first-order term in the Lax \emph{et al.} expansion \cite{EandS}.

By following the same route that lead us to (\ref{Eparax3}), it is not difficult to see that within the same approximations (\ref{Mono&Parax})  the positive-frequency part of the magnetic field operator can be written as
\begin{align}
 \hvB^+ & \! (\vvr,t)=  \frac{i}{c}  \int_0^\infty \di \omega \left( {\frac{\hbar \omega}{4 \pi \varepsilon_0 c}}\right)^{1/2} \exp \bigl[-i \omega(t-z/c) \bigr] \nonumber \\
  & \! \! \! \! \! \! \! \! \!\times  \sum_{\mu,n,m} \ha_{\mu nm}(\omega)\left( \hz \times \hat{\bm{x}}_\mu +i \frac{c}{\omega} \hx_\mu \times \bm{\nabla}_\perp \right) \psi_{nm}(\vx,z, \omega), \label{Bparax3}
\end{align}
where the symbol ``$\times$''  denotes the ordinary vector cross product.
\subsection{Monochromatic limit}
In the laboratory practice, one often deals with laser beams whose bandwidth is so narrow that they can be considered basically monochromatic. For this case the formalism that we have developed above may be redundant and simplified expressions can be used. In order to pass from the general case above to the monochromatic limit it is convenient to make, as a preliminary step, the passage from a continuous to a discrete frequency spectrum
 by letting
\begin{align}
\omega \; \rightarrow \; \omega_j = \omega_0 + j \Delta \omega, \label{OmDiscr1}
\end{align}
where $j \in \{ 0, \pm1, \pm2, \dots \}$. In this limit the continuous-frequency annihilation operators are transformed to the discrete-frequency ones via the rule
\begin{align}
\ha_{\mu nm}(\omega) \; \rightarrow \; \left(\Delta \omega \right)^{-1/2} \ha_{\mu nm j}, \label{OmDiscr2}
\end{align}
and integrals over continuous frequency are converted to sums over the discrete index $j$ according to
\begin{align}
\int \di \omega \; \rightarrow \; \Delta \omega  \sum_j\;, \qquad \delta(\omega - \omega') \; \rightarrow \; (\Delta \omega)^{-1} \delta_{j j'}. \label{OmDiscr3}
\end{align}
The discrete-frequency electric field operator is obtained by applying the rules (\ref{OmDiscr1}-\ref{OmDiscr3}) in (\ref{Eparax3}), thus obtaining
\begin{align}
\hvE^+ & \! (\vvr,t) =  i  \sum_{j} \left({\frac{\hbar \omega_j }{2 \varepsilon_0 c/\Delta \omega}}\right)^{1/2}\exp \left[-i \omega_j(t-z/c) \right] \nonumber \\
&  \times \sum_{\mu,  n,m}   \ha_{\mu nm j}  \left( \hat{\bm{x}}_\mu +i \hz \frac{c}{\omega_j} \hx_\mu \cdot \bm{\nabla}_\perp \right)\psi_{nm}(\vx,z, \omega_j), \label{Eparax4}
\end{align}
where the discrete-frequency annihilation operator $\ha_{\mu n m j}$ satisfies the commutation rules
\begin{align}
\bigl[\ha_{\mu nm j},\ha_{\mu' n' m' j'}^\dagger\bigr] = \delta_{\mu \mu'} \delta_{nn'} \delta_{mm'} \delta_{j j'}. \label{CCR5}
\end{align}
For excitations of the radiation field that satisfy (\ref{Mono&Parax}) we can keep the solely term $j = 0$ in (\ref{Eparax4}) to obtain the strict monochromatic limit
\begin{align}
\hvE^+ & \! (\vvr,t) =  i  \left({\frac{\hbar \omega_0 }{2 \varepsilon_0 c/\Delta \omega}}\right)^{1/2} \exp \left[-i \omega_0(t-z/c) \right] \nonumber \\
&  \times \sum_{\mu,n,m}  \ha_{\mu nm}  \left( \hat{\bm{x}}_\mu +i \hz \frac{c}{\omega_0} \hx_\mu \cdot \bm{\nabla}_\perp \right)\psi_{nm}(\vx,z, \omega_0), \label{Eparax5}
\end{align}
where we used the shorthand $\ha_{\mu nm} \equiv \left. \ha_{\mu nm j}\right|_{j=0}$.
It is worth noting that in the expression above, the strict paraxial limit would be obtained only by neglecting the term proportional to $\hx_\mu \cdot \bm{\nabla}_\perp $.

The same procedure that lead to Eq. (\ref{Eparax5}) can be followed to obtain the following expression for the positive-frequency part of the magnetic field operator in the monochromatic limit:
\begin{align}
\hvB^+ & \! (\vvr,t) =  \frac{i}{c}  \left({\frac{\hbar \omega_0 }{2 \varepsilon_0 c/\Delta \omega}}\right)^{1/2} \exp \left[-i \omega_0(t-z/c) \right] \nonumber \\
  & \! \! \! \! \! \! \! \! \! \times \sum_{\mu,n,m}  \ha_{\mu nm}  \left( \hz \times \hat{\bm{x}}_\mu +i  \frac{c}{\omega_0} \hx_\mu \times \bm{\nabla}_\perp \right)\psi_{nm}(\vx,z, \omega_0). \label{Bparax5}
\end{align}
\section{Linear and angular momentum of the field}
In the previous section we have derived explicit expressions for the electric and magnetic field operators. This allows us now to calculate both the linear and the angular momentum of the quantized electromagnetic field.
\subsection{Linear momentum}
The normal-order linear momentum density operator $:\! \hvP(\vvr,t)\! :$ is equal to $1/c^2$ the Poynting vector normal-order operator $:\! \hvS(\vvr,t)\! :$ and it is expressed in terms of the electric and magnetic field operators  as \cite{MandelBook}:
\begin{align}
:\! \hvP(\vvr,t)\!:\; = \frac{\varepsilon_0}{2}  \left[ : \hvE(\vvr,t) \times \hvB(\vvr,t) -  \hvB(\vvr,t) \times \hvE(\vvr,t) : \right] .\label{LinMom0}
\end{align}
If in the expression above we substitute
\begin{align}
\hvE(\vvr,t) & =   \hvE^+(\vvr,t) + \hvE^-(\vvr,t), \label{E0} \\
 \hvB(\vvr,t) & =  \hvB^+(\vvr,t) + \hvB^-(\vvr,t),  \label{B0}
\end{align}
with
\begin{align}
 \hvE^-(\vvr,t) =  \bigl[ \hvE^+(\vvr,t) \bigr]^\dagger , \quad
 \hvB^-(\vvr,t) = \bigl[ \hvB^+(\vvr,t) \bigr]^\dagger,  \label{Bpm}
\end{align}
we can write Eq. (\ref{LinMom0}) as the sum of $8$ terms:
\begin{align}
:\!\hvP(\vvr,t)\!: \;=  &  \; \frac{\varepsilon_0}{2} :\!\Bigl\{ \left( \hvE^- \times \hvB^+ -  \hvB^- \times \hvE^+\right)  \nonumber \\
& \hphantom{ ii}  + \left( \hvE^+ \times \hvB^- - \hvB^+ \times \hvE^-\right) \nonumber
\nonumber \\
& \hphantom{ ii}    + \left( \hvE^- \times \hvB^- -  \hvB^- \times \hvE^-\right)  \nonumber \\
& \hphantom{ ii}    + \left( \hvE^+ \times \hvB^+ - \hvB^+ \times \hvE^+\right) \Bigr\} \!: \; , \label{LinMom1}
\end{align}
where, for sake of clarity, we have omitted the explicit space and time dependence of the electric and magnetic field operators. The last four addenda of this sum contain integrands with terms as $\hat{a}_{\mu n m}(\omega)\hat{a}_{\mu' n' m'}(\omega')$ and $\hat{a}_{\mu n m}^\dagger(\omega)\hat{a}_{\mu' n' m'}^\dagger(\omega')$ which oscillates at frequencies higher than $\omega : \omega_0 - \Delta \omega \leq \omega \leq \omega_0 + \Delta \omega$ where, once again, $\omega_0$ is the central frequency of the beam and $\Delta \omega$ its bandwidth. As most optical detectors integrate the received signal over a time interval $T$ much longer than $1/\Delta \omega$, the last four terms in Eq. (\ref{LinMom1}) can be neglected in the limit $T \gg 1/\Delta \omega$. Within this assumption, the time-integrated power flow trough the detector surface (coincident with the $x,y$ plane) or, equivalently,  the linear momentum operator per unit length, can be written as:
\begin{align}
\hat{\bm{P}} =\; \, : \iint\limits_{-\infty}^{\quad  \infty} \left( \frac{1}{T} \int_{-T/2}^{T/2} \hvP(\vvr,t) \, \di t  \right) \di x \, \di y  : \;,\label{PspaceAv}
\end{align}
where the spatial integration is extended over the whole $\mathbb{R}^2$ plane. Although
the integration time $T$ in Eq. (\ref{PspaceAv}) is finite, the integration interval $[-T/2,T/2]$ can be formally estended to $(-\infty, \infty)$ without significant error for narrow-band beams.
Thus, substitution in Eq. (\ref{PspaceAv}) from Eqs. (\ref{Eparax3}-\ref{Bparax3}) gives
\begin{align}
 \hat{\bm{P}} \, =&\,  \frac{1}{c^2 \, T}  \int_0^\infty \di \omega \,\hbar \omega \biggl\{ \sum_{\mu=1}^2 \sum_{n,m} \sum_{n',m'} \bm{\mathcal{P}}^{nm,n'm'} \nonumber \\
&   \times \ha_{\mu n m}^\dagger(\omega) \ha_{\mu n' m'}(\omega) \biggr\}, \label{eq10}
\end{align}
where
\begin{align}
\bm{\mathcal{P}}^{nm,n'm'}& \equiv \iint\limits_{-\infty}^{\quad  \infty} \psi^*_{nm}\left(\hvz - i \lbar \bm{\nabla}_\perp \right) \psi_{n'm'} \, \di x \di y \nonumber \\
& =  \frac{\theta_0}{2 \, i} \biggl[  \hvx \, \delta_{m m'} \Bigl( \sqrt{n'} \, \delta_{n', n+1} -\sqrt{n} \, \delta_{n, n'+1} \Bigr) \nonumber \\
& \hphantom{\frac{\theta_0}{2 i} \Bigl[ } + \,
 \hvy \, \delta_{n n'} \Bigl( \sqrt{m'} \, \delta_{m', m+1} -\sqrt{m} \, \delta_{m, m'+1} \Bigr) \biggr] \nonumber \\
& \hphantom{\frac{\theta_0}{2 i} \Bigl[ } + \,
  \hvz \, \delta_{n n'}  \delta_{m m'} , \label{eq20}
\end{align}
with $\lbar = 2 \pi c/ \omega$.
Not surprisingly, Eq. (\ref{eq10}) has the same form of Eq. (21) in Ref. \cite{vanEnk&Nienhuis94}. The main difference is in the form of the quasi-paraxial linear momentum operator $\hvz - i \lbar \bm{\nabla}_\perp$ as compared with the truly ``quantum-mechanical'' linear momentum operator $- i \hbar \bm{\nabla}$ in coordinate representation. This difference is substantial and will manifest its effects later, when we will calculate orbital angular momentum operators and their commutation relations. The strict paraxial limit is obtained by tacking the limit $\theta_0 \rightarrow 0$ in Eq. (\ref{eq10}). In this case only the longitudinal component $\mathcal{P}_z$ keeps a nonzero value, but this is inconsistent with a correct representation of the energy flow in the beam \cite{EandS}.

Finally, the three components of the operator $ \hat{\bm{P}}$  can be calculated explicitly by substituting Eq. (\ref{eq20}) into Eq. (\ref{eq10}) obtaining
\begin{align}
 &    \hat{P}_x \, =  \frac{1}{c^2 \, T}  \frac{\theta_0}{2 \, i} \int_0^\infty \di \omega \,\hbar \omega \biggl\{ \sum_{\mu=1}^2 \sum_{n,m}  (n+1)^{1/2} \nonumber \\
&   \times    \Bigl[ \ha_{\mu n m}^\dagger(\omega) \ha_{\mu, n+1, m}(\omega)   - \ha_{\mu, n+1, m}^\dagger(\omega) \ha_{\mu n m}(\omega) \Bigr] \biggr\},
\end{align}
\begin{align}
 &  \hat{P}_y\, =  \frac{1}{ c^2 \,T}  \frac{\theta_0}{2 \, i}  \int_0^\infty \di \omega \,\hbar \omega \biggl\{ \sum_{\mu=1}^2 \sum_{n,m}  (m+1)^{1/2} \nonumber \\
 & \times  \Bigl[ \ha_{\mu n m}^\dagger(\omega) \ha_{\mu n, m+1}(\omega)   - \ha_{\mu n, m+1}^\dagger(\omega) \ha_{\mu n m}(\omega) \Bigr] \biggr\},
\end{align}
for the transverse components, and
\begin{align}
    \hat{P}_z \, =&\,  \frac{1}{c^2 \, T}  \int_0^\infty \di \omega \,\hbar \omega \sum_{\mu=1}^2 \sum_{n,m}  \ha_{\mu n m}^\dagger(\omega) \ha_{\mu n m}(\omega),
\end{align}
for the longitudinal one, which is simply proportional to the total number of photons per unit length of the beam.
\medskip
\subsection{Angular momentum}
The calculation of the normal-order angular momentum density operator $:\!\hvJ(\vvr,t)\!:$ proceeds along the line delineated in the previous subsection, starting form the standard definition
\begin{align}
:\! \hvJ(\vvr,t) \! : \; = \; :\! \vvr \times \hvP(\vvr,t) \! : \;, \label{Jdensity}
\end{align}
and arriving to the time-integrated angular momentum operator per unit length:
\begin{align}
\hat{\bm{J}} = \; :  \iint\limits_{-\infty}^{\quad  \infty} \left( \frac{1}{T} \int_{-T/2}^{T/2} \hvJ(\vvr,t) \, \di t  \right) \, \di x \di y  : \;. \label{JspaceAv}
\end{align}
Again, substitution in Eq. (\ref{JspaceAv}) from Eqs. (\ref{Eparax3}-\ref{Bparax3}) gives
\begin{align}
 \hat{\bm{J}} \, =&\,  \frac{1}{c^2 \, T}  \int_0^\infty \di \omega \,\hbar \omega \biggl\{ \sum_{\mu,\mu'} \sum_{n,m} \sum_{n',m'} \bm{\mathcal{J}}^{\mu nm, \,\mu 'n'm'} \nonumber \\
&   \times \ha_{\mu n m}^\dagger(\omega) \ha_{\mu' n' m'}(\omega) \biggr\}, \label{JspaceAv2}
\end{align}
where we have defined
\begin{widetext}
\begin{align}
\bm{\mathcal{J}}^{\mu nm, \,\mu 'n'm'} \equiv & \, \int \psi^*_{nm}\left[-i \lbar \left( \hvx_\mu \times \hvx_{\mu'} \right)  + \delta_{\mu \mu'} \, \vvr \times \left( \hvz -i \lbar  \bm{\nabla}_\perp  \right)
\right] \psi_{n'm'}  \, \di x  \di y \nonumber \\
= & \, \delta_{\mu \mu'} \frac{\lbar}{\theta_0}  \biggl\{ \hvx \, \delta_{n n'} \left[ m^{1/2} \delta_{m,m'+1} + (m')^{1/2} \delta_{m',m+1} \right]  - \hvy \, \delta_{m m'} \left[ n^{1/2} \delta_{n,n'+1} + (n')^{1/2} \delta_{n',n+1} \right] \biggr\}
 \nonumber \\
&
 -i \lbar \hvz \biggl\{\sigma_{\mu \mu'} \delta_{n n'} \delta_{m m'} + \delta_{\mu \mu'} \left[ (n m')^{1/2} \delta_{n,n'+1}\delta_{m',m+1}  -(n' m)^{1/2} \delta_{n',n+1}\delta_{m,m'+1}
 \right]
 \biggr\}, \label{eq30}
\end{align}
\end{widetext}
with $\hvx_\mu \times \hvx_{\mu'}  = \hvz \varepsilon_{\mu \mu' 3} \equiv \hvz \sigma_{\mu \mu'} $.  From Eqs. (\ref{eq20},\ref{eq30}) it follows that we can write
\begin{align}
\bm{\mathcal{J}}= \bm{\mathcal{S}} \otimes \bm{\mathcal{I}}_L + \bm{\mathcal{I}}_S \otimes \bm{\mathcal{L}}  , \label{eq40}
\end{align}
where $\bm{\mathcal{I}}_{S}$ and  $\bm{\mathcal{I}}_{L}$  are the identity operators in the spin and orbital angular momentum spaces, respectively, and we have defined
\begin{align}
\left( \bm{\mathcal{S}} \cdot \hvz \right)_{\mu \mu'} & \, = - i  \lbar \varepsilon_{\mu \mu' 3}  ,\label{eq50} \\
\bm{\mathcal{L}} & \, = \vvr \times \bm{\mathcal{P}}, \label{eq60}
\end{align}
with  $\mu,\mu' \in \{ 1,2\}$,  and $\bm{\mathcal{S}} \cdot \hvx = 0 = \bm{\mathcal{S}} \cdot \hvy$.
This result is the quasi-paraxial analogous of Eq. (20) of Ref. \cite{vanEnk&Nienhuis94}, and it shows the separation of the total angular momentum of the beam in its spin and orbital parts. In the paraxial limit $\theta_0 \rightarrow 0$, Eq. (\ref{eq30}) is dominated by the transverse part and  both the spin and the angular contribution to $\hat{J}_z$ become negligible. This is in agreement with the results of Haus and Pan \cite{HausandPan} who have shown that a self-consistent description of angular momentum of light beams cannot be achieved in a purely paraxial context.

From Eq. (\ref{eq30}) the three components of the angular momentum operator per unit length can be explicitly calculated obtaining
\begin{align}
 & \, \hat{J}_x =  \frac{1}{c^2 \, T}  \frac{\lbar}{\theta_0}   \int_0^\infty \di \omega \,\hbar \omega \biggl\{ \sum_{\mu=1}^2 \sum_{n,m}  (m+1)^{1/2} \nonumber \\
&   \times  \Bigl[ \ha_{\mu n m}^\dagger(\omega) \ha_{\mu n, m+1}(\omega)  + \ha_{\mu n, m+1}^\dagger(\omega) \ha_{\mu n m}(\omega) \Bigr] \biggr\},
\end{align}
\begin{align}
 & \hat{J}_y =  -\frac{1}{c^2 \, T}   \frac{\lbar}{\theta_0}  \int_0^\infty \di \omega \,\hbar \omega \biggl\{ \sum_{\mu=1}^2 \sum_{n,m}  (n+1)^{1/2} \nonumber \\
&   \times \Bigl[ \ha_{\mu n m}^\dagger(\omega) \ha_{\mu, n+1, m}(\omega)   + \ha_{\mu, n+1, m}^\dagger(\omega) \ha_{\mu n m}(\omega) \Bigr] \biggr\},
\end{align}
and $\hat{J}_z \equiv \hat{S}_z + \hat{L}_z$, where we have defined the spin and the orbital angular momentum longitudinal components, respectively as
\begin{align}
\hat{S}_z= & \,  \frac{ 1}{c^2 \, T} \frac{\lbar}{i} \int_0^\infty \di \omega \,\hbar \omega   \sum_{n,m}\left[ \ha_{1 n m}^\dagger(\omega) \ha_{2 n m}(\omega) \right. \nonumber \\
 & \,\left. - \ha_{2 n m}^\dagger(\omega) \ha_{1 n m}(\omega)\right], \label{JzS}
\end{align}
and
\begin{align}
& \hat{L}_z =  \frac{1}{c^2 \, T} \frac{\lbar}{i} \int_0^\infty \di \omega \,\hbar \omega    \sum_{\mu,n,m}\left[(n+1) (m+1)\right]^{1/2} \nonumber \\
& \times \left[ \ha_{\mu, n +1,m}^\dagger(\omega) \ha_{\mu n, m+1}(\omega)  - \ha_{\mu n, m+1}^\dagger(\omega) \ha_{\mu, n+1, m}(\omega)\right]. \label{JzL}
\end{align}
\section{Commutation rules}
At this point we have collected all the ingredients necessary to calculate the commutation relations between the linear and angular momentum operator components. From Eqs. (\ref{eq10},\ref{JspaceAv2}) it follows that both these operators  have the form:
\begin{align}
\hat{U}_a & \, =   \frac{1}{c^2 \, T}  \int_0^\infty \di \omega \,  \hbar \omega  \sum_{A,A'} \ha_{A}^\dagger  \, \mathcal{U}_a^{A A'}  \ha_{A'}, \label{OpForm}
\end{align}
where $U \in \{ J, P\}$, $\mathcal{U} \in \{ \mathcal{J}, \mathcal{P}\}$, and $a \in \{ x,y,z\}$. In addition we have introduced the cumulative labels $A; A'; \ldots$ that embody the three indices $\mu, n, m;\; \mu', n', m';\ldots$ respectively, so that, e.g., $\sum_A = \sum_\mu \sum_n \sum_n$. Now, we assume the validity of the following commutation relations:
\begin{align}
\bigl[ {\mathcal{U}}_a, {\mathcal{V}}_b \bigr] =  i \lbar f_{abc}^{\mathcal{U}\mathcal{V}} \mathcal{W}_c, \label{CCR1}
\end{align}
where $U,V,W \in \{ J, P\}$, $\mathcal{U},\mathcal{V},\mathcal{W} \in \{ \mathcal{J}, \mathcal{P}\}$, $a,b,c \in \{ x,y,z\}$, and $f_{abc}^{\mathcal{U}\mathcal{V}} $ are numerical coefficient to be determined and summation over repeated indices is understood. Then, it is not difficult to see that
\begin{align}
  \bigl[ \hat{U}_a, \hat{V}_b \bigr] & \, =   \frac{1}{c^4 \, T^2}  \int_0^\infty \di \omega \, \left( \hbar \omega \right)^2 \sum_{A,A'} \bigl[ {\mathcal{U}}_a^{A A'}, {\mathcal{V}}_b^{A A'} \bigr] \ha_{A}^\dagger  \ha_{A'} \nonumber \\
& \, =   \frac{i \hbar}{c \, T} \left(\frac{f_{abc}^{\mathcal{U}\mathcal{V}}}{c^2 \, T}  \int_0^\infty \di \omega \,  \hbar \omega \sum_{A,A'} \ha_{A}^\dagger  {\mathcal{W}}_c^{A A'} \ha_{A'} \right) \nonumber \\
& \, =  \frac{ i \hbar}{c \, T} f_{abc}^{\mathcal{U}\mathcal{V}} \,\hat{W}_c, \label{CCR3}
\end{align}
where the factor $c T$ in the last line of the equation above, plays the role of ``natural'' unit length for the problem under consideration. We remind that the validity of Eq. (\ref{CCR3}) is subject to the  assumption (\ref{CCR1}) that must be still verified. We begin such check by noticing that from Eq. (\ref{eq20})  it is easy to calculate
\begin{align}
  \bigl[ \mathcal{P}_a, \mathcal{P}_b \bigr] & \, = 0, \qquad a,b \in \{ x,y,z\} ,
\end{align}
which has the form (\ref{CCR1}) with $f_{abc}^{\mathcal{P}\mathcal{P}} =0$. Similarly,
from Eqs. (\ref{eq30},\ref{eq40}) it follows that
\begin{align}
  \bigl[ \mathcal{J}_a, \mathcal{J}_b \bigr] & \, = \mathcal{I} \otimes   \bigl[ \mathcal{L}_a, \mathcal{L}_b \bigr] +  \bigl[ \mathcal{S}_a, \mathcal{S}_b \bigr] \otimes  \mathcal{I}. \label{eq70}
\end{align}
which reduces to its first term solely since ${\mathcal{S}_x =0= \mathcal{S}_y}$ trivially implies
\begin{align}
  \bigl[ \mathcal{S}_a, \mathcal{S}_b \bigr] & \, = 0, \qquad a,b \in \{ x,y,z\} .
\end{align}
By using Eq. (\ref{eq30}), an explicit calculation furnishes
\begin{align}
  \bigl[ \mathcal{L}_x, \mathcal{L}_y \bigr] & \, = 0, \label{CCRL1} \\
  \bigl[ \mathcal{L}_x, \mathcal{L}_z \bigr] & \, = - i \lbar \mathcal{L}_y, \label{CCRL2} \\
  \bigl[ \mathcal{L}_y, \mathcal{L}_z \bigr] & \, =  i \lbar \mathcal{L}_x , \label{CCRL3}
\end{align}
that, together with Eq. (\ref{eq70}) gives $f_{abc}^{\mathcal{J}\mathcal{J}} = \varepsilon_{abc}(1 - \delta_{cz})$, that amounts to a violation of canonical commutation relations for the total angular momentum. Before discussing this somewhat surprising results, let us conclude the calculations by showing that
\begin{align}
  \bigl[ \mathcal{L}_a, \mathcal{P}_z \bigr] & \, = 0, \label{CCRP1} \\
  \bigl[ \mathcal{L}_a, \mathcal{P}_a \bigr] & \, = 0, \label{CCRP2} \\
  \bigl[ \mathcal{L}_x, \mathcal{P}_y \bigr] & \, =  i \lbar \mathcal{P}_z , \label{CCRP3} \\
  \bigl[ \mathcal{L}_y, \mathcal{P}_x \bigr] & \, =  -i \lbar \mathcal{P}_z  \label{CCRP4}\\
  \bigl[ \mathcal{L}_z, \mathcal{P}_x \bigr] & \, =  i \lbar \mathcal{P}_y  \label{CCRP5} \\
  \bigl[ \mathcal{L}_z, \mathcal{P}_y \bigr] & \, =  -i \lbar \mathcal{P}_x  \label{CCRP6},
\end{align}
which follows from Eqs. (\ref{eq20},\ref{eq30}-\ref{eq40}) and from the trivial identity $\bigl[ \mathcal{P}_a, \mathcal{S}_b \bigr]=0$, as $\bm{\mathcal{P}}$ and $\bm{\mathcal{S}}$ operates upon different linear spaces. For short, Eqs. (\ref{CCRP1}-\ref{CCRP6}) give  ${f_{abc}^{\mathcal{L}\mathcal{P}} = \varepsilon_{abc}(1 -  \delta_{bz})}$.
\subsection{Discussion}
In order to discuss the results present above, it is useful to adopt the notation $\check{O}$ to indicate the operator $\check{O}$ in non-relativistic two-dimensional quantum mechanics in the Schr\"{o}dinger picture. Thus, for example,  $\check{\vx} = \vx$, $\check{\vp} = -i \hbar \, \bm{\nabla}_\perp$, etc. Note that in this representation the longitudinal coordinate $z$ is \emph{not} a dynamical variable, but a parameter that plays the role of ``time'' in the  Schr\"{o}dinger equation \cite{PhysRevA.48.656}.
If with $| n \rangle,\,| n' \rangle, \dots$ we denote the eigenstates  of a one-dimensional harmonic oscillator, then from Eq. (\ref{eq20}) it follows that
\begin{align}
\mathcal{P}_x \propto & \,\langle n | \check{p}_x | n' \rangle\langle m | m' \rangle, \label{HO1} \\
\mathcal{P}_y \propto & \,\langle n |  n' \rangle\langle m |\check{p}_y | m' \rangle, \label{HO2} \\
\mathcal{P}_z \propto & \,\langle n |  n' \rangle\langle m | m' \rangle. \label{HO3}
\end{align}
 The equations above show that while $\mathcal{P}_x$ and $\mathcal{P}_y$ behaves like Cartesian components of the canonical linear momentum, $\mathcal{P}_z$ does not as it is proportional to the identity operator.
 In a similar manner we can see that Eq. (\ref{eq30}) implies
\begin{align}
\mathcal{L}_x \propto & \,\langle n | \check{y} | n' \rangle\langle m | m' \rangle, \label{HO4} \\
\mathcal{L}_y \propto & \,- \langle n |  n' \rangle\langle m |\check{x} | m' \rangle, \label{HO5} \\
\mathcal{L}_z \propto & \,\langle n,m |\check{x}\check{p}_y - \check{y}\check{p}_x|  n' , m' \rangle, \label{HO6}
\end{align}
namely $\mathcal{L}_x \sim \check{y}$, $\mathcal{L}_y \sim -\check{x}$ and $\mathcal{L}_z \sim \check{L}_z$, where the symbol ``$ \sim $'' stands for ``behaves like'', and $\check{L}_z = \check{x} \check{p}_y - \check{y} \check{p}_x$.
If we look at Eqs. (\ref{HO1}-\ref{HO6}) the origin of the anomalous commutation relations (\ref{CCRL1}-\ref{CCRP6}) becomes clear, since
\begin{align}
  \bigl[ \mathcal{L}_x, \mathcal{L}_y \bigr] & \, \sim \bigl[ \check{y}, -\check{x} \bigr] =0, \label{CCRL1a} \\
  \bigl[ \mathcal{L}_x, \mathcal{L}_z \bigr] & \,\sim \bigl[ \check{y}, \check{L}_z \bigr] = i \hbar \check{x} \sim  - i \lbar \mathcal{L}_y, \label{CCRL2a} \\
  \bigl[ \mathcal{L}_x, \mathcal{P}_y \bigr] & \,\sim \bigl[ \check{y}, \check{p}_y \bigr] = i \hbar  \sim   i \lbar \mathcal{P}_z, \label{CCRL3a}
\end{align}
and so on.
In physical terms, the essence is that any well collimated beam is basically an eigenstate of the $z$-component of the linear momentum operator which, for such beams, practically reduces to a $c$-number. Thus,
in order to recover canonical commutation relations, it is necessary to deal with beams with either high angular aperture $\theta_0$ for whose our first-order approximation breaks down, or with a direction of propagation that deviates from the reference axis $z$ by an angle grater than $\theta_0$ \cite{aiello:100401}. As a final remark, it should be noticed that violations of canonical commutation relations for the angular momentum of an electromagnetic field of arbitrary shape, were already reported by van Enk and Nienhuis \cite{vanEnk&Nienhuis94}.
\section{Examples}
In the previous section we have completed the study of the formal properties of the linear and angular momentum operators for quasi-paraxial beams. In this section we will illustrate the usefulness of our treatment by applying it to the common case of coherent excitations of the electromagnetic field.

Let us begin by considering the quantum coherent state representing an ordinary monochromatic laser beam prepared on the laboratory bench in the spatial mode $\psi_{nm}(\vx,\omega_0)$ and linearly polarized along the direction $\hvx_\mu$:
\begin{align}
|\alpha \rangle = \exp \left( \alpha \hat{a}^\dagger_A - \alpha^* \hat{a}_A \right) | 0\rangle,
\end{align}
where $A = \mu, n, m$. For non-monochromatic coherent states of the field,  this simple expression immediately generalize to \cite{PhysRevA.42.4102}
\begin{align}
| [\alpha_{A}] \rangle  = \hat{D}[\alpha_{A}]| 0 \rangle,
\end{align}
where
\begin{align}
\hat{D}[\alpha_{A}] = \exp \bigg\{  \int_0^\infty \di \omega \Big(
 \alpha_{A}(\omega) \hat{a}^\dagger_{A}(\omega) - \alpha_{A}^*(\omega) \hat{a}_{A}(\omega)
 \Big) \bigg\}.
\end{align}
is the displacement operator for the multi-mode quasi-paraxial electromagnetic field, and square brackets $[\alpha_{A}]$ indicate functional dependence. It is easy to see via a direct calculation that the amplitude functions $\alpha_{A}(\omega)$ are the eigenvalues of the annihilation operator   $\hat{a}_{A} (\omega) $:
\begin{align}
\hat{a}_{A} (\omega) | [\alpha_{A'}] \rangle = \alpha_A(\omega) \delta_{A A'} | [\alpha_{A}] \rangle,
\end{align}
where $\delta_{A A'} = \delta_{\mu \mu'} \delta_{n n'} \delta_{m m'}$.
The coherent state $| [\alpha_{A}] \rangle$ is an eigenstate of the positive-frequency part of the electric field operator \cite{DeutschAJP}
\begin{align}
\hat{\vE}^+ (\vvr,t) | [\alpha_{A}] \rangle = {\vE}^+_\text{cl}[\alpha_A] (\vvr,t) | [\alpha_{A}] \rangle, \label{eigenfield}
\end{align}
where the eigenvalue $\vE^+_\text{cl}[\alpha_A](\vvr,t)$ is the analytic signal \cite{MandelBook} of the  classical field generated by the excitation with spectral amplitude $\alpha_A = \alpha_{\mu nm}(\omega)$ of the mode $\psi_{nm}$:
\begin{align}
\vE^+_\text{cl}[\alpha_A] & \!(\vvr,t)=  i  \int_0^\infty \di \omega \left( {\frac{\hbar \omega}{4 \pi \varepsilon_0 c}}\right)^{1/2} \exp \bigl[-i \omega(t-z/c) \bigr] \nonumber \\
&  \times  \alpha_{\mu nm}(\omega)\left( \hat{\bm{x}}_\mu +i \hz \frac{c}{\omega} \hx_\mu \cdot \bm{\nabla}_\perp \right)\psi_{nm}(\vx,z, \omega).
\end{align}
Multi-mode field coherent excitations are handled exactly in the same manner by writing the $N$-mode coherent state $| \Psi_N \rangle = | \{ [\alpha_{A}] \} \rangle$ as
\begin{align}
| \Psi_N \rangle = & \,| [\alpha_{A_1}] \rangle | [\alpha_{A_2}] \rangle \cdot \ldots \cdot | [\alpha_{A_N}] \rangle = \prod_{i = 1}^N | [\alpha_{A_i}] \rangle, \label{eq100}
\end{align}
and Eq. (\ref{eigenfield}) becomes
\begin{align}
\hat{\vE}^+ (\vvr,t) | \Psi_N \rangle = {\vE}^+_\text{cl} [\{ \alpha_A \}](\vvr,t) | \Psi_N \rangle, \label{eq110}
\end{align}
where
\begin{align}
\vE^+_\text{cl} & \![\{ \alpha_A \}] (\vvr,t) \nonumber \\
& =  \sum_{i = 1}^N \vE^+_\text{cl} [ \alpha_{A_i} ](\vvr,t)  \nonumber  \\
 & =  i  \int_0^\infty \di \omega \left( {\frac{\hbar \omega}{4 \pi \varepsilon_0 c}}\right)^{1/2} \exp \bigl[-i \omega(t-z/c) \bigr] \nonumber \\
&  \times  \sum_{\mu,n,m}\alpha_{\mu nm}(\omega)\left( \hat{\bm{x}}_\mu +i \hz \frac{c}{\omega} \hx_\mu \cdot \bm{\nabla}_\perp \right)\psi_{nm}(\vx,z, \omega). \label{eq120}
\end{align}

Equations (\ref{eq100}-\ref{eq120}) describe the more general coherent state excitation for a quasi-paraxial beam. The expectation value of linear and angular momentum operators with respect to these states can be written, after a straightforward calculation, as
\begin{align}
\langle \Psi_N | \hat{\bm{U}} | \Psi_N \rangle & \,=  \frac{1}{c^2 T} \int_0^\infty \di \omega
\hbar \omega \sum_{i,j}^{1,N} \alpha^*_{A_i}(\omega) \bm{\mathcal{U}}^{A_i A_j} \alpha_{A_j} (\omega) \nonumber \\
& \,=  \frac{1}{c^2 T} \int_0^\infty \di \omega
\hbar \omega  \bigl( \vec{\alpha}(\omega) ,\bm{\mathcal{U}}\, \vec{\alpha}(\omega) \bigr)_N, \label{Uav}
\end{align}
where $U\in \{ J, P\}$, $\mathcal{U} \in \{ \mathcal{J}, \mathcal{P}\}$, and we have used the suggestive notation $\bigl( \vec{\alpha}(\omega) , \vec{\beta}(\omega) \bigr)_N$ to indicate the scalar product between the two $N$-dimensional vectors $\vec{\alpha}(\omega) = \left( \alpha_{A_1}, \ldots, \alpha_{A_N} \right)$ and $\vec{\beta}(\omega) = \left( \beta_{A_1}, \ldots, \beta_{A_N} \right)$.
For a monochromatic beam of central frequency $\omega_0$, Eq. (\ref{OmDiscr2}) requires $\alpha_A(\omega) \rightarrow (\Delta \omega)^{-1/2} \alpha_A^D(\omega_0) $ (the superscript $D$ stands for ``discrete''), and Eq. (\ref{Uav}) reduces to
\begin{align}
\langle \Psi_N | \hat{\bm{U}} | \Psi_N \rangle = \frac{\hbar \omega_0}{c^2 T}  \bigl( \vec{\alpha}^D(\omega_0) ,\bm{\mathcal{U}}\, \vec{\alpha}^D(\omega_0) \bigr)_N. \label{Uav0}
\end{align}

It is instructive to apply Eq. (\ref{Uav0}) to a concrete case in order to see the physical meaning of the equations above. With this aim, let us consider a spatial mode of the field of the form $\hat{\mathbf{u}} f(\vx,z,\omega_0)$, where $\hat{\mathbf{u}} = \xi \hat{\mathbf{x}} + \eta \hat{\mathbf{y}}$, with $\abs{\xi}^2 + \abs{\eta}^2 =1$, is a unit vector that fixes the polarization  of the beam, and
\begin{align}
f(\vx,z,\omega_0) = \alpha_{00} \psi_{00} +  \alpha_{10} \psi_{10}+ \alpha_{01} \psi_{01} ,
\end{align}
where  $\psi_{nm} = \psi_{nm}(\vx,z,\omega_0)$ indicates the $(n,m)$ Hermite-Gaussian mode \cite{SiegmanBook}, and $\abs{\alpha_{00}}^2 + \abs{\alpha_{10}}^2 +  \abs{\alpha_{01}}^2=1$ because of normalization.   The corresponding coherent excitation $ | \Psi_{N=6} \rangle$ can be written as
\begin{align}
 | \Psi_6 \rangle = & \, | [\alpha_{100}] \rangle| [\alpha_{110}] \rangle| [\alpha_{101}] \rangle | [\alpha_{200}] \rangle| [\alpha_{210}] \rangle| [\alpha_{201}] \rangle\\ \nonumber
= & \, |\xi \alpha_{00} \rangle| \xi \alpha_{10} \rangle| \xi \alpha_{01} \rangle|\eta \alpha_{00} \rangle| \eta \alpha_{10} \rangle| \eta \alpha_{01} \rangle,
\end{align}
where we have defined $\alpha_{1nm}(\omega_0) \equiv \alpha_{nm}$. Substitution in Eq. (\ref{Uav0}) from Eq. (\ref{eq20}) gives
\begin{align}
\langle \Psi_6 | \hat{\bm{P}} | \Psi_6 \rangle = \frac{\hbar \omega_0}{c^2 T}  \left( \hbx P_x + \hby P_y + \hbz P_z\right),
\end{align}
with
\begin{align}
P_x = & \, \theta_0 \, \text{Im} \left(\alpha_{00}^* \alpha_{10} \right)  , \label{avP1} \\
P_y = & \, \theta_0 \, \text{Im} \left(\alpha_{00}^* \alpha_{01} \right)   , \label{avP2} \\
P_z = & \,  1. \label{avP3}
\end{align}
The expressions above are rich of information.  First, Eq. (\ref{avP3}) shows that $P_z= \abs{\alpha_{00}}^2 + \abs{\alpha_{10}}^2 +  \abs{\alpha_{01}}^2=1$ furnishes  the total intensity of the beam, and that it is unbiased with respect to the modes  $\psi_{nm}$, thus revealing both its ``identity'' character. Second, if we define $\tan \theta_\mu \equiv P_\mu/P_z, \; (\mu \in \{x,y\})$, then from Eqs. (\ref{avP1}-\ref{avP2}) it follows that the spatial mode $f(\vx,z,\omega_0)$ deviates from the axis $z$ only when $\alpha_{00} \neq 0$, namely only when the fundamental Gaussian mode $\psi_{00}$ is present in such superposition. In other words, in order to define an ``absolute deviation'' it is necessary the presence of a reference mode $\psi_{nn}$ which is symmetric with respect to the simultaneous inversion $x \rightarrow -x, \, y \rightarrow - y$. Finally, when $\alpha_{00} \in \mathbb{R}$, Eqs. (\ref{avP1}-\ref{avP2}) reproduce the well-known result that either $\alpha_{10}$ or $\alpha_{01}$ must have an imaginary part to guarantee a nonzero tilting angle $\theta_\mu$ \cite{TrepsTilt}.

In a similar manner we can evaluate the expectation value of the angular momentum operator obtaining
\begin{align}
\langle \Psi_6 | \hat{\bm{J}} | \Psi_6 \rangle = \frac{\hbar \omega_0}{c^2 T}  \left( \hbx J_x + \hby J_y + \hbz J_z\right)
\end{align}
where
\begin{align}
J_x = & \, w_0 \, \text{Re} \left(\alpha_{00}^* \alpha_{01} \right)  , \label{avJ1} \\
J_y = & \,  - w_0 \, \text{Re} \left(\alpha_{00}^* \alpha_{10} \right)   , \label{avJ2} \\
J_z = & \, \lbar  \left[\, \sigma + 2 \, \text{Im} \left(\alpha_{10}^* \alpha_{01} \right) \right] , \label{avJ3}
\end{align}
with $\sigma = i (\xi \eta^* - \xi^* \eta)$ denoting the \emph{elicity} of the beam.
Here, if $\alpha_{00} \in \mathbb{R}$, Eqs. (\ref{avJ1}) and (\ref{avJ2}) imply that in order to have a nonzero transverse orbital angular momentum either $\alpha_{10}$ or $\alpha_{01}$ must have a \emph{real} part. However, it is  well known that a superposition with real coefficients of the fundamental mode $\psi_{00}$ with either $\psi_{10}$ or $\psi_{01}$  describes approximatively a \emph{displaced} Gaussian beam along the $x$- or the $y$-axis, respectively \cite{TrepsTilt}. In other words, a lateral displacement of a Gaussian beam changes its transverse angular momentum or, vice versa, the occurrence of non zero transverse components of the angular momentum cause a transverse displacement of the beam \cite{aiello:100401}. On the other hand, it is also known that a transverse displacement cannot affect the longitudinal angular momentum $J_z$ \cite{BerrySPIE,VasneTilt}, as it is confirmed by Eq. (\ref{avJ3}) which goes to zero when both $\alpha_{10}$ and $\alpha_{01}$ are real numbers.
However, for a \emph{pure} Laguerre-Gaussian beam we have $f(\vx,z,\omega_0) = \psi^\text{LG}_{\ell=\pm 1, p=0} = (\psi_{10} \pm i \psi_{01})/\sqrt{2}$, namely $\alpha_{00} =0 \Rightarrow J_x =0 = J_y$.
Thus,  in this case $2 \, \text{Im} \left( \alpha_{10}^* \alpha_{01}\right) = \pm 1$ and Eq. (\ref{avJ3}) furnishes $J_z = \lbar ( \sigma \pm 1)$ which, in agreement with previous calculations \cite{Allen92}, shows that a Laguerre-Gaussian beam possesses $\abs{\ell}= 1$ units of orbital angular momentum along of the direction of propagation.

It worth noting that the results of this section for a coherent beam are in perfect agreement with the classical results presented in Ref. \cite{aiello:100401}.
\section{Summary}
In this paper we have applied the theory of quantized light beams to investigate the properties of the transverse components of the  angular momentum operator of the electromagnetic field. It is known that for either massive particles or photons localized in wave packets it is meaningful to talk about the \emph{total} angular momentum of the system under consideration. This is evaluated as the integral over the whole $\mathbb{R}^3$ space, of the angular momentum \emph{density} of either the particle or of the field. However, when dealing with photons in beam-like states, the relevant quantity which can be actually measured on a laboratory bench, is the angular momentum per unit length. This is evaluated, in a plane of equation $z=$ const.  perpendicular to the main direction of propagation $z$ of the beam, as the integral over the transverse $xy$-plane  of the angular momentum {density} of the field. Since in this case the integration does not extend over all $3$D space, the angular momentum per unit length is not independent of time \cite{vanEnk&Nienhuis}. Therefore, it becomes necessary to average this quantity over the measurement time $T$, and one is led to the expression shown in Eq. (\ref{JspaceAv}). We have explicitly evaluated this time-averaged angular momentum per unit length $\hat{\bm{J}}$ within the framework of paraxial optics, but including contributions from first-order transverse derivatives of the electric and magnetic fields. This inclusion permitted us to achieve a self-consistent description of both the energy flow (linear momentum or Poynting vector) and the spin and orbital angular momentum of the beam which appear to be naturally separated in the sum $\hat{\bm{J}} =\hat{\bm{S}}  + \hat{\bm{L}}$. Then, we have calculated the commutation relations between the Cartesian components of $\hat{\bm{J}},\hat{\bm{L}},\hat{\bm{S}}$ and we have found that they differ from the standard one. In particular, while $\hat{L}_z$ is still a bona fide generator of rotations around the propagation axis $z$, the transverse components  $\hat{L}_x$ and  $\hat{L}_y$ commute and, as it was already found at a classical level \cite{aiello:100401}, they are strictly connected with the transverse coordinates $y$ and $-x$ of center of the beam, respectively. Finally, as a realistic example illustrating the above mentioned connection, we calculated the expectation value of $\hat{\bm{J}}$ between multi-mode coherent states of the electromagnetic field. It is well known that for these states quantum and classical expectation values basically coincide.  Indeed, we found full consistency between our results and the classical ones \cite{aiello:100401}, namely we found that $\langle \hat{L}_x \rangle \propto \Delta_y$ and $\langle \hat{L}_y \rangle \propto -\Delta_x$, where  $\Delta_x$ and $\Delta_y$ are the transverse displacements of the center of the beam with respect to the propagation axis $z$. Further investigations of the relations between angular momentum and beam shifts in classical optics, are illustrated in Refs. \cite{AielloOL08,Bliokh:09} and references therein.
\section{Acknowledgements}
AA
acknowledges support from the Alexander von Humboldt
Foundation.
%

\begin{thebibliography}{28}
\expandafter\ifx\csname natexlab\endcsname\relax\def\natexlab#1{#1}\fi
\expandafter\ifx\csname bibnamefont\endcsname\relax
  \def\bibnamefont#1{#1}\fi
\expandafter\ifx\csname bibfnamefont\endcsname\relax
  \def\bibfnamefont#1{#1}\fi
\expandafter\ifx\csname citenamefont\endcsname\relax
  \def\citenamefont#1{#1}\fi
\expandafter\ifx\csname url\endcsname\relax
  \def\url#1{\texttt{#1}}\fi
\expandafter\ifx\csname urlprefix\endcsname\relax\def\urlprefix{URL }\fi
\providecommand{\bibinfo}[2]{#2}
\providecommand{\eprint}[2][]{\url{#2}}

\bibitem[{\citenamefont{Allen et~al.}(1992)\citenamefont{Allen, Beijersbergen,
  Spreeuw, and Woerdman}}]{Allen92}
\bibinfo{author}{\bibfnamefont{L.}~\bibnamefont{Allen}},
  \bibinfo{author}{\bibfnamefont{M.~W.} \bibnamefont{Beijersbergen}},
  \bibinfo{author}{\bibfnamefont{R.~J.~C.} \bibnamefont{Spreeuw}},
  \bibnamefont{and} \bibinfo{author}{\bibfnamefont{J.~P.}
  \bibnamefont{Woerdman}}, \bibinfo{journal}{Phys. Rev. A}
  \textbf{\bibinfo{volume}{45}}, \bibinfo{pages}{8185} (\bibinfo{year}{1992}).

\bibitem[{\citenamefont{Siegman}(1986)}]{SiegmanBook}
\bibinfo{author}{\bibfnamefont{A.~E.} \bibnamefont{Siegman}},
  \emph{\bibinfo{title}{Lasers}} (\bibinfo{publisher}{University Science
  Books}, \bibinfo{address}{Mill Valley, CA}, \bibinfo{year}{1986}).

\bibitem[{Not()}]{NotePhoton}
\bibinfo{note}{For the intense beams of light considered in \cite{Allen92} with
  ``number of photons per unit of volume $N/V$'' we mean the ratio between the
  energy density $W$ of the beam and the single-photon energy $\hbar \omega_0$
  at the central frequency $\omega_0$ of the beam: $N/V = W/(\hbar \omega_0)$.}

\bibitem[{\citenamefont{Sasada and Okamoto}(2003)}]{PhysRevA.68.012323}
\bibinfo{author}{\bibfnamefont{H.}~\bibnamefont{Sasada}} \bibnamefont{and}
  \bibinfo{author}{\bibfnamefont{M.}~\bibnamefont{Okamoto}},
  \bibinfo{journal}{Phys. Rev. A} \textbf{\bibinfo{volume}{68}},
  \bibinfo{pages}{012323} (\bibinfo{year}{2003}).

\bibitem[{\citenamefont{Hsu et~al.}(2009)\citenamefont{Hsu, Bowen, and
  Lam}}]{hsu:043825}
\bibinfo{author}{\bibfnamefont{M.~T.~L.} \bibnamefont{Hsu}},
  \bibinfo{author}{\bibfnamefont{W.~P.} \bibnamefont{Bowen}}, \bibnamefont{and}
  \bibinfo{author}{\bibfnamefont{P.~K.} \bibnamefont{Lam}},
  \bibinfo{journal}{Phys. Rev. A} \textbf{\bibinfo{volume}{79}},
  \bibinfo{eid}{043825} (\bibinfo{year}{2009}).

\bibitem[{\citenamefont{Allen et~al.}(2003)\citenamefont{Allen, Barnett, and
  Padgett}}]{BarnettBook}
\bibinfo{editor}{\bibfnamefont{L.}~\bibnamefont{Allen}},
  \bibinfo{editor}{\bibfnamefont{S.~M.} \bibnamefont{Barnett}},
  \bibnamefont{and} \bibinfo{editor}{\bibfnamefont{M.~J.}
  \bibnamefont{Padgett}}, eds., \emph{\bibinfo{title}{Optical Angular
  Momentum}} (\bibinfo{publisher}{Institute of Physics Publishing},
  \bibinfo{address}{Bristol, UK}, \bibinfo{year}{2003}).

\bibitem[{\citenamefont{Franke-Arnold et~al.}(2008)\citenamefont{Franke-Arnold,
  Allen, and Padgett}}]{FrankeA}
\bibinfo{author}{\bibfnamefont{S.}~\bibnamefont{Franke-Arnold}},
  \bibinfo{author}{\bibfnamefont{L.}~\bibnamefont{Allen}}, \bibnamefont{and}
  \bibinfo{author}{\bibfnamefont{M.}~\bibnamefont{Padgett}},
  \bibinfo{journal}{Laser \& Photon. Rev.} \textbf{\bibinfo{volume}{2}},
  \bibinfo{pages}{299} (\bibinfo{year}{2008}).

\bibitem[{\citenamefont{{van Enk} and Nienhuis}(1992)}]{vanEnk&Nienhuis}
\bibinfo{author}{\bibfnamefont{S.~J.} \bibnamefont{{van Enk}}}
  \bibnamefont{and} \bibinfo{author}{\bibfnamefont{G.}~\bibnamefont{Nienhuis}},
  \bibinfo{journal}{Opt. Commun.} \textbf{\bibinfo{volume}{94}},
  \bibinfo{pages}{147} (\bibinfo{year}{1992}).

\bibitem[{\citenamefont{Barnett and Allen}(1994)}]{Barnett&Allen}
\bibinfo{author}{\bibfnamefont{S.~M.} \bibnamefont{Barnett}} \bibnamefont{and}
  \bibinfo{author}{\bibfnamefont{L.}~\bibnamefont{Allen}},
  \bibinfo{journal}{Opt. Commun.} \textbf{\bibinfo{volume}{110}},
  \bibinfo{pages}{670} (\bibinfo{year}{1994}).

\bibitem[{\citenamefont{{van Enk} and Nienhuis}(1994)}]{vanEnk&Nienhuis94}
\bibinfo{author}{\bibfnamefont{S.~J.} \bibnamefont{{van Enk}}}
  \bibnamefont{and} \bibinfo{author}{\bibfnamefont{G.}~\bibnamefont{Nienhuis}},
  \bibinfo{journal}{J. Mod. Opt.} \textbf{\bibinfo{volume}{41}},
  \bibinfo{pages}{963} (\bibinfo{year}{1994}).

\bibitem[{\citenamefont{Calvo et~al.}(2006)\citenamefont{Calvo, Pic\'{o}n, and
  Bagan}}]{calvo:013805}
\bibinfo{author}{\bibfnamefont{G.~F.} \bibnamefont{Calvo}},
  \bibinfo{author}{\bibfnamefont{A.}~\bibnamefont{Pic\'{o}n}},
  \bibnamefont{and} \bibinfo{author}{\bibfnamefont{E.}~\bibnamefont{Bagan}},
  \bibinfo{journal}{Phys. Rev. A} \textbf{\bibinfo{volume}{73}},
  \bibinfo{eid}{013805} (\bibinfo{year}{2006}).

\bibitem[{\citenamefont{Aiello et~al.}(2009)\citenamefont{Aiello, Lindlein,
  Marquardt, and Leuchs}}]{aiello:100401}
\bibinfo{author}{\bibfnamefont{A.}~\bibnamefont{Aiello}},
  \bibinfo{author}{\bibfnamefont{N.}~\bibnamefont{Lindlein}},
  \bibinfo{author}{\bibfnamefont{C.}~\bibnamefont{Marquardt}},
  \bibnamefont{and} \bibinfo{author}{\bibfnamefont{G.}~\bibnamefont{Leuchs}},
  \bibinfo{journal}{Phys. Rev. Lett.} \textbf{\bibinfo{volume}{103}},
  \bibinfo{eid}{100401} (\bibinfo{year}{2009}).

\bibitem[{\citenamefont{Bekshaev}(2009)}]{Bekshaev:09}
\bibinfo{author}{\bibfnamefont{A.~Y.} \bibnamefont{Bekshaev}},
  \bibinfo{journal}{J. Opt. A: Pure Appl. Opt.} \textbf{\bibinfo{volume}{11}},
  \bibinfo{pages}{094003} (\bibinfo{year}{2009}).

\bibitem[{\citenamefont{Aiello and Woerdman}(2005)}]{AielloPRAParax}
\bibinfo{author}{\bibfnamefont{A.}~\bibnamefont{Aiello}} \bibnamefont{and}
  \bibinfo{author}{\bibfnamefont{J.~P.} \bibnamefont{Woerdman}},
  \bibinfo{journal}{Phys. Rev. A} \textbf{\bibinfo{volume}{72}},
  \bibinfo{pages}{060101(R)} (\bibinfo{year}{2005}).

\bibitem[{\citenamefont{Aiello et~al.}(2006)\citenamefont{Aiello, Visser,
  Nienhuis, and Woerdman}}]{AielloOLParax}
\bibinfo{author}{\bibfnamefont{A.}~\bibnamefont{Aiello}},
  \bibinfo{author}{\bibfnamefont{J.}~\bibnamefont{Visser}},
  \bibinfo{author}{\bibfnamefont{G.}~\bibnamefont{Nienhuis}}, \bibnamefont{and}
  \bibinfo{author}{\bibfnamefont{J.~P.} \bibnamefont{Woerdman}},
  \bibinfo{journal}{Opt. Lett.} \textbf{\bibinfo{volume}{31}},
  \bibinfo{pages}{525} (\bibinfo{year}{2006}).

\bibitem[{\citenamefont{Lax et~al.}(1975)\citenamefont{Lax, Louisell, and
  McKnight}}]{PhysRevA.11.1365}
\bibinfo{author}{\bibfnamefont{M.}~\bibnamefont{Lax}},
  \bibinfo{author}{\bibfnamefont{W.~H.} \bibnamefont{Louisell}},
  \bibnamefont{and} \bibinfo{author}{\bibfnamefont{W.~B.}
  \bibnamefont{McKnight}}, \bibinfo{journal}{Phys. Rev. A}
  \textbf{\bibinfo{volume}{11}}, \bibinfo{pages}{1365} (\bibinfo{year}{1975}).

\bibitem[{\citenamefont{Haus and Pan}(1993)}]{HausandPan}
\bibinfo{author}{\bibfnamefont{H.~A.} \bibnamefont{Haus}} \bibnamefont{and}
  \bibinfo{author}{\bibfnamefont{J.~L.} \bibnamefont{Pan}},
  \bibinfo{journal}{Am. J. Phys.} \textbf{\bibinfo{volume}{61}},
  \bibinfo{pages}{818} (\bibinfo{year}{1993}).

\bibitem[{\citenamefont{Erikson and Singh}(1994)}]{EandS}
\bibinfo{author}{\bibfnamefont{W.~L.} \bibnamefont{Erikson}} \bibnamefont{and}
  \bibinfo{author}{\bibfnamefont{S.}~\bibnamefont{Singh}},
  \bibinfo{journal}{Phys. Rev. E} \textbf{\bibinfo{volume}{49}},
  \bibinfo{pages}{5778} (\bibinfo{year}{1994}).

\bibitem[{Rod()}]{Rodrigues}
\bibinfo{note}{S. Belongie. ``Rodrigues' Rotation Formula.'' From MathWorld--A
  Wolfram Web Resource, created by Eric W. Weisstein.
  http://mathworld.wolfram.com/... \\/RodriguesRotationFormula.html}.

\bibitem[{\citenamefont{Mandel and Wolf}(1995)}]{MandelBook}
\bibinfo{author}{\bibfnamefont{L.}~\bibnamefont{Mandel}} \bibnamefont{and}
  \bibinfo{author}{\bibfnamefont{E.}~\bibnamefont{Wolf}},
  \emph{\bibinfo{title}{Optical coherence and quantum optics}}
  (\bibinfo{publisher}{Cambridge University Press},
  \bibinfo{address}{Cambridge, UK}, \bibinfo{year}{1995}),
  \bibinfo{edition}{1st} ed.

\bibitem[{\citenamefont{Nienhuis and Allen}(1993)}]{PhysRevA.48.656}
\bibinfo{author}{\bibfnamefont{G.}~\bibnamefont{Nienhuis}} \bibnamefont{and}
  \bibinfo{author}{\bibfnamefont{L.}~\bibnamefont{Allen}},
  \bibinfo{journal}{Phys. Rev. A} \textbf{\bibinfo{volume}{48}},
  \bibinfo{pages}{656} (\bibinfo{year}{1993}).

\bibitem[{\citenamefont{Blow et~al.}(1990)\citenamefont{Blow, Loudon, Phoenix,
  and Shepherd}}]{PhysRevA.42.4102}
\bibinfo{author}{\bibfnamefont{K.~J.} \bibnamefont{Blow}},
  \bibinfo{author}{\bibfnamefont{R.}~\bibnamefont{Loudon}},
  \bibinfo{author}{\bibfnamefont{S.~J.~D.} \bibnamefont{Phoenix}},
  \bibnamefont{and} \bibinfo{author}{\bibfnamefont{T.~J.}
  \bibnamefont{Shepherd}}, \bibinfo{journal}{Phys. Rev. A}
  \textbf{\bibinfo{volume}{42}}, \bibinfo{pages}{4102} (\bibinfo{year}{1990}).

\bibitem[{\citenamefont{Deutsch}(1991)}]{DeutschAJP}
\bibinfo{author}{\bibfnamefont{I.~H.} \bibnamefont{Deutsch}},
  \bibinfo{journal}{Am. J. Phys.} \textbf{\bibinfo{volume}{59}},
  \bibinfo{pages}{834} (\bibinfo{year}{1991}).

\bibitem[{\citenamefont{Hsu et~al.}(2005)\citenamefont{Hsu, Bowen, Treps, and
  Lam}}]{TrepsTilt}
\bibinfo{author}{\bibfnamefont{M.~T.~L.} \bibnamefont{Hsu}},
  \bibinfo{author}{\bibfnamefont{W.~P.} \bibnamefont{Bowen}},
  \bibinfo{author}{\bibfnamefont{N.}~\bibnamefont{Treps}}, \bibnamefont{and}
  \bibinfo{author}{\bibfnamefont{P.~K.} \bibnamefont{Lam}},
  \bibinfo{journal}{Phys. Rev. A} \textbf{\bibinfo{volume}{72}},
  \bibinfo{pages}{013802} (\bibinfo{year}{2005}).

\bibitem[{\citenamefont{Berry}(1998)}]{BerrySPIE}
\bibinfo{author}{\bibfnamefont{M.}~\bibnamefont{Berry}},
  \bibinfo{journal}{Proc. SPIE (Int. Conf. on Singular Optics)}
  \textbf{\bibinfo{volume}{3487}}, \bibinfo{pages}{6} (\bibinfo{year}{1998}).

\bibitem[{\citenamefont{Vasnetsov et~al.}(2005)\citenamefont{Vasnetsov,
  {Pas'ko}, and Soskin}}]{VasneTilt}
\bibinfo{author}{\bibfnamefont{M.~V.} \bibnamefont{Vasnetsov}},
  \bibinfo{author}{\bibfnamefont{V.~A.} \bibnamefont{{Pas'ko}}},
  \bibnamefont{and} \bibinfo{author}{\bibfnamefont{M.~S.}
  \bibnamefont{Soskin}}, \bibinfo{journal}{New Journal of Physics}
  \textbf{\bibinfo{volume}{7}}, \bibinfo{pages}{46} (\bibinfo{year}{2005}).

\bibitem[{\citenamefont{Aiello and Woerdman}(2008)}]{AielloOL08}
\bibinfo{author}{\bibfnamefont{A.}~\bibnamefont{Aiello}} \bibnamefont{and}
  \bibinfo{author}{\bibfnamefont{J.~P.} \bibnamefont{Woerdman}},
  \bibinfo{journal}{Opt. Lett.} \textbf{\bibinfo{volume}{33}},
  \bibinfo{pages}{1437} (\bibinfo{year}{2008}).

\bibitem[{\citenamefont{Bliokh et~al.}(2009)\citenamefont{Bliokh, Shadrivov,
  and Kivshar}}]{Bliokh:09}
\bibinfo{author}{\bibfnamefont{K.~Y.} \bibnamefont{Bliokh}},
  \bibinfo{author}{\bibfnamefont{I.~V.} \bibnamefont{Shadrivov}},
  \bibnamefont{and} \bibinfo{author}{\bibfnamefont{Y.~S.}
  \bibnamefont{Kivshar}}, \bibinfo{journal}{Opt. Lett.}
  \textbf{\bibinfo{volume}{34}}, \bibinfo{pages}{389} (\bibinfo{year}{2009}).

\end{thebibliography}
%

\end{document}